\documentclass{webofc}
\usepackage[varg]{txfonts}   
\usepackage{hyperref}
\newcommand{\mm}{(\mu^+\mu^-)}

\begin{document}

\title{Predicting and Discovering True Muonium}

\author{
\firstname{Henry} \lastname{Lamm}\inst{1}\fnsep\thanks{\email{hlamm@umd.edu}} \and
\firstname{Yao} \lastname{Ji}\inst{2}\fnsep\thanks{\email{yao.ji@physik.uni-regensburg.de}}
        }

\institute{Department of Physics, University of Maryland, College Park, MD 20742, USA 
\and
Institut f\"ur Theoretische Physik, Universit\"at Regensburg, Regensburg 93040, Germany}

\abstract{The recent observation of discrepancies in the muonic sector motivates searches for the yet undiscovered atom true muonium $\mm$. To leverage potential experimental signals, precise theoretical calculations are required. I will present the on-going work to compute higher-order corrections to the hyperfine splitting and the Lamb shift. Further, possible detection in rare meson decay experiments like REDTOP and using true muonium production to constrain mesonic form factors will be discussed.
}
\maketitle
\section{Introduction}
Given the dearth of clear signals beyond the Standard Model (BSM) from the LHC, it may prove useful to consider more subtle deviations from well-understood observables.  Under this paradigm, one might organize future searches around resolving the ``muon problem'': the curious coincidence that multiple observables in the muonic sector deviate from either theoretical predictions or similar results with other leptonic flavors.  

The most persistent deviation is the the anomalous magnetic moment $a_\mu$ measured at Brookhaven~\cite{PhysRevD.73.072003} to deviate by $\approx3\sigma$ from the theoretical predictions.  Upcoming experiments at Fermilab~\cite{Grange:2015fou} and J-PARC~\cite{Saito:2012zz} are expected to reduce the experimental uncertainty by a factor of four.  Simultaneously, the largest two theoretical uncertainties, the hadronic vacuum polarization and hadronic light-by-light, are expected to be reduced sufficiently that if the current mean values persist, the discrepancy would exceed 5$\sigma$.  Another long-standing discrepancy in the low energy sector, the charge radii from muonic atoms~\cite{Antognini:1900ns,Pohl1:2016xoo}, appears to be resolving itself with electronic measurements~\cite{Beyer79} with near-term experiments to clarify this issue further.  If rectified in favor of the muonic results, these observables will put stringent constraints on new physics. At higher energies, the ratio of leptonic decays in $D$ and $B$ mesons have found $2-4\sigma$ discrepancies with expectations~\cite{Aaij:2014ora,Aaij:2015yra,Ciezarek:2017yzh,Aaij:2017tyk,Aaij:2017vbb}. 

A new class of observables that can shed light on the muon problem are those associated with the bound state $(\mu^+\mu^-)$~\cite{TuckerSmith:2010ra,PhysRevD.91.073008,Lamm:2015gka,Lamm:2016jim}.  This state has alternatively been dubbed ``true muonium''~\cite{Hughes:1971}, ``bimuonium''~\cite{baier1962bimuonium}, and ``dimuonium''~\cite{budini1961reactions}.  Simpler bound states like positronium $(e^+e^-)$, hydrogen, and muonium $(\mu^+e^-)$ have attracted significant attention as testing grounds for precision QED studies~\cite{Karshenboim:2005iy}, but are limited in their BSM discovery potential by either the mass suppression $\mathcal{O}(m_e/\Lambda_{BSM})$ or large theoretical uncertainties from unknown nuclear structure effects.  In contrast, true muonium has a much larger reduced mass ($\mu=\frac{m_\mu}{2}$), and its QCD corrections are limited to the better-understood hadronic loop effects.

Alas, true muonium has yet to be directly observed.  The first reason is that it is experimentally difficult to producing low-energy muon pairs, and is exacerbated by the bound state's short lifetime ($\tau\approx$ 1 ps), which presents an interesting challenge to experimenters.  A second, more prosaic, reason for the neglect is until the $a_\mu$ anomaly, it seemed unlikely true muonium would offer any novel physics justifying the large effort.

In this talk, we present state of the art theoretical predictions for key energy splittings and lifetimes.  Following this, we discuss the possibilities to observe true muonium in upcoming experiments.

\section{Predicting}
Resolutions to the muon problem relying upon BSM generically lead to $\mathcal{O}(100 $ MHz) corrections to transitions and decay rates of true muonium (e.g, \cite{TuckerSmith:2010ra}). These $10^{-6}-10^{-9}$ corrections are of plausible size for measurement if the production of large numbers of atoms were possible. By virtue of the annihilation channel, true muonium observables (e.g. transitions, production and decay rates) are sensitive at lower order to new particle content than other atomic transitions where individual measurements can be insensitive to different particle content (i.e. Pseudoscalar contributions to the Lamb shift in $(\mu H)$ are heavily suppressed).  $\mathcal{O}(100 $ MHz) is estimated be $\mathcal{O}(m\alpha^7)$ in true muonium; therefore, our goal should be to predictions of this level, where hadronic and electroweak effects must be taken into account\cite{Lamm:2016vtf,PhysRevD.91.073008}  

The theoretical expression for the energy levels to true muonium from QED can 
be written
\begin{align}
\label{eq:eng}
 E_{n,l,j,s}=-\frac{m_\mu\alpha^2}{4n^2}+m_\mu\alpha^4\bigg[&C_0+C_1\frac{\alpha}{\pi}+C_{21}\alpha^2\ln\left(\frac{
1}{\alpha}\right)+C_{20}\left(\frac{\alpha}{\pi}\right)^2\nonumber\\
 &+C_{32}\frac{\alpha^3}{\pi}\ln^2\left(\frac{1}{\alpha}\right)+C_{31}\frac{
\alpha^3}{\pi}\ln\left(\frac{1}{\alpha}\right)+C_{30}\left(\frac{\alpha}{\pi}\right)^3+\cdots\bigg],
\end{align}
where $C_{ij}$ indicate the coefficient of the term proportional to 
$(\alpha)^i\ln^j(1/\alpha)$.  $C_{ij}$ include any dependence on mass scales 
other than $m_\mu$.  The coefficients of 
single-flavor QED bound 
states, used in positronium, are known up to $\mathcal{O}(m_e\alpha^6)$. 
Partial results exist for $\mathcal{O}(m_e\alpha^7)$ and are an 
active research area (For updated reviews of 
the coefficients see \cite{Adkins:2014dva,PhysRevA.94.032507}).

True muonium has extra contributions that must be considered.  Large $C_{ij}$ arise for electronic contributions due to $m_{\mu}/m_e\approx200$.  In fact, the Lamb shift is dominated by the electronic vacuum polarization~\cite{Jentschura:1997tv}.  The relative smallness of $m_\tau/m_\mu\approx 17$ and $m_{\pi}/m_{\mu}\approx1.3$ produces contributions to true muonium much larger than analogous contributions to positronium.

Since \cite{Lamm:2015fia}, a number of important contributions have been computed and an update of several important transitions are shown in Table~\ref{tab:1}.  For the HFS, improved calculations of the $\mathcal{O}(m_\mu\alpha^5)$ contributions and large $\mathcal{O}(m_\mu\alpha^6)$ and $\mathcal{O}(m_\mu\alpha^7)$ have reduced the uncertainty be a factor-of-4~\cite{PhysRevA.94.032507,Adkins:2015jia,Lamm:2016vtf,Lamm:2017lib}.  
Spin-independent contributions to the spectra are only known partially at $\mathcal{O}(m_\mu\alpha^5)$, but some large $\mathcal{O}(m_\mu\alpha^6)$ terms have been computed and included the presented results~\cite{Naveen}.

\begin{table}[h]
\begin{center}
\caption{\label{tab:1}Theoretical predictions for key transitions in true muonium.  The \textit{had}-lab 
electron allows for large loop contributionseled uncertainty corresponds to the hadronic contributions, while unlabeled uncertainties indicated estimates of missing contributions}
 \begin{tabular}{c c}
 \hline\hline
 Transition& $E_{\rm theory}~\rm[MHz]$\\
 \hline
  $1^3S_1-1^1S_0$&$42329355(51)_{had}(700)$\\
  $2^3S_1-1^3S_1$&$2.550014(16)\times10^{11}$\\
  $2^3P_0-2^3S_1$&$1.002(3)\times10^7$\\
  $2^3P_1-2^3S_1$&$1.115(3)\times10^7$\\
  $2^3P_2-2^3S_1$&$1.206(3)\times10^7$\\
  $2^1P_1-2^3S_1$&$1.153(3)\times10^7$\\
  \hline\hline
 \end{tabular}
\end{center}
\end{table}

It should be emphasized that the missing contributions require no new theoretical techniques; positronium, muonium, and muonic hydrogen techniques can be straight-forwardly applied.  Most of the unknown corrections arise from virtual electron loops to photon propagators.

Due in part to the difficulty of in-beam laser spectroscopy, consideration of other methods of measuring the Lamb shifts should be considered.  An older method utilized for atomic hydrogen~\cite{Robiscoe:1965zz} has been suggested for true muonium~\cite{Brodsky:2009gx} which is similar to the methods pursued currently by the DIRAC experiment for ($\pi^+\pi^-$)~\cite{Nemenov:2001vp}.  In this method, a beam of $2S$ state true muonium would be passed through a magnetic field, resulting in level-mixing with $2P$ state which decays by $\gamma$ emission to the $1S$, decreasing the intensity of the beam.  By measuring the beam's intensity as a function of magnetic field, the Lamb shift can be extracted.

Since $m_\mu>m_e$, it is possible for $n^3S_1$ states to decay into $e^+e^-$ pairs.  All currently contemplated searches utilize this decay for discovering true muonium~\cite{Celentano:2014wya,dirac,Bogomyagkov:2017uul}, so predictions of these rates are desirable.  Including all NLO and a large NNLO contribution we find,
\begin{align}
\Gamma(1^3S_1\rightarrow e^+e^-)&=\left(1+\left[-\frac{221}{36}+\frac{4}{3}\ln\left(2\frac{m_\mu}{m_e}\right)-0.3899(8)\right]\frac{\alpha}{\pi}+455(1)\frac{\alpha^2}{\pi^2}\right)\frac{\alpha^5m_\mu}{6}\\\nonumber
&\implies\tau(1^3S_1\rightarrow e^+e^-)=\frac{1}{\Gamma(1^3S_1\rightarrow e^+e^-)}=1.79560(13)\times10^{-12}\text{ s},\\
 \Gamma(2^3S_1\rightarrow e^+e^-)&=\left(1+\left[-\frac{221}{36}+\frac{4}{3}\ln\left(2\frac{m_\mu}{m_e}\right)-0.3899(8)\right]\frac{\alpha}{\pi}+394(1)\frac{\alpha^2}{\pi^2}\right)\frac{\alpha^5m_\mu}{48}\\\nonumber
&\implies\tau(2^3S_1\rightarrow e^+e^-)=\frac{1}{\Gamma(2^3S_1\rightarrow e^+e^-)}=14.3696(10)\times10^{-12}\text{ s},\\
 \Gamma(3^3S_1\rightarrow e^+e^-)&=\left(1+\left[-\frac{221}{36}+\frac{4}{3}\ln\left(2\frac{m_\mu}{m_e}\right)-0.3899(8)\right]\frac{\alpha}{\pi}+400(100)\frac{\alpha^2}{\pi^2}\right)\frac{\alpha^5m_\mu}{162}\\\nonumber
&\implies\tau(3^3S_1\rightarrow e^+e^-)=\frac{1}{\Gamma(3^3S_1\rightarrow e^+e^-)}=48.5(3)\times10^{-12}\text{ s},\\
\end{align}
where the final NLO term comes from hadronic vacuum polarization~\cite{Lamm:2016vtf} and its associated error, and the NNLO coefficient's error is estimated by $\mathcal{O}(1)$ for the $n=1,2$ because other contributions are not anticipated to be anomalously large. For $n=3$ the large error is because even the anomalously large NNLO contribution is not known so we estimate based on the lower $n$ and assign a 25\% uncertainty.  Even at this precision, the theoretical values are lower than the 1\%, 5\%, and 15\% respectively for $n=1,2,3$ that has been suggested as experimental uncertainties attainable at a near-term experiment~\cite{Bogomyagkov:2017uul}.

Singlet states will predominately decay to two photons, and similar precision is known for these rates~\cite{Jentschura:1997tv}, but we quote here the leading order values $\tau(n^1S_0\rightarrow\gamma\gamma)=0.6n^3\times10^{-12}$ s.  

If very high-intensity true muonium experiments were ever built, it would be possible to measure more exotic decays, including those of the triplet state to neutrinos.  The leading order decay rates to mono-energetic neutrinos are known to be $\Gamma(1^3S_1\rightarrow\nu_\mu\bar{\nu}_\mu)\approx10^{-11}\Gamma_{e^+e^-}$ and $\Gamma(1^3S_1\rightarrow \nu_l\bar{\nu}_l)\approx10^{-14}\Gamma_{e^+e^-}$.  These rates are admittedly small but unlike positronium, due to a $\propto m_\ell^5$ scaling, are around the level of rare mesonic decays.  Further,  measurement of neutrino decays are related to the general subject of invisible decays.  These were shown to constrain a variety of BSM (e.g. extra dimensions, axions, mirror matter, fractional charges, and other low-mass dark matter models) in positronium~\cite{Badertscher:2006fm}.  In true muonium, these rates are also enhanced due to mass scaling, as well as raising the upper limit on masses from $m_e$ to $m_\mu$.

\section{Discovering}
The greatest experimental obstacle is producing a sufficiently large number of the bound state.  In the past, many production channels have been discussed:  $\pi p\rightarrow (\mu^+\mu^-)n$~\cite{Bilenky:1969zd}, $\gamma Z\rightarrow (\mu^+\mu^-) Z$~\cite{Bilenky:1969zd}, $e Z\rightarrow e(\mu^+\mu^-) Z$~\cite{Holvik:1986ty,ArteagaRomero:2000yh},  $\mu^+\mu^-\rightarrow (\mu^+\mu^-)$~\cite{Hughes:1971}, $e^+e^-\rightarrow (\mu^+\mu^-)$~\cite{baier1961bimuonium,Moffat:1975uw,Brodsky:2009gx}, $e^+e^-\rightarrow (\mu^+\mu^-)\gamma$\cite{Brodsky:2009gx}, $\eta\rightarrow (\mu^+\mu^-)\gamma$~\cite{Nemenov:1972ph,Kozlov:1987ey}, $K_L\rightarrow (\mu^+\mu^-)\gamma$~\cite{Malenfant:1987tm}, $Z_1Z_2\rightarrow Z_1Z_2(\mu^+\mu^-)$~\cite{Ginzburg:1998df}, and $q^+q^-\rightarrow(\mu^+\mu^-)g$ in a quark plasma~\cite{Chen:2012ci}.  Some of the more novel methods of utilizing these production channels considered include: fixed target experiments~\cite{Banburski:2012tk}, Fool's Intersection Storage Rings~\cite{Brodsky:2009gx}, and even from astrophysical sources~\cite{Ellis:2015eea,Ellis:2017kzh}.

Two fixed-target experiments should be highlighted in this discussion.  The Heavy Photon Search (HPS)~\cite{Celentano:2014wya} experiment has plans to search for true muonium~\cite{Banburski:2012tk}, and DImeson Relativistic Atom Complex (DIRAC)~\cite{Benelli:2012bw} has discussed the possibility in an upgraded run~\cite{dirac}.  Additionally, the DIRAC experiment intends to study the Lamb shift in the $(\pi^+\pi^-)$ bound state using a fixed magnetic field and measuring the decay rate as a function of distance~\cite{Nemenov:2001vp}, and the methods developed could be applied to true muonium.  

Brodsky and Lebed's idea to produce true muonium by colliding $e^+e^-$ at an angle with energy near $2m_\mu$~\cite{Brodsky:2009gx} has been developed by a group at BINP into an idea for a low-energy collider~\cite{Bogomyagkov:2017uul}.  Their proposed physics analysis beyond discovering true muonium could include: production rates, decay lengths, $2P-1S$ transition probabilities, $2P$ lifetimes, and possibly even transition energies. 

Another avenue for discovering true muonium is through a rare meson decay.  The proposed REDTOP experiment at Fermilab would search for true muonium in $\eta/\eta'$ decays~\cite{Gatto:2016rae}.  In anticipation of this, the rates have been recomputed using modern form factors derived from dispersive techniques and applying NLO corrections that will be explained in an upcoming work:
\begin{align}
 &\frac{\mathcal{B}(\eta\rightarrow\gamma(\mu^+\mu^-))}{\mathcal{B}(\eta\rightarrow\gamma\gamma)}=1.476(5)_{\rm stat}(4)_{\rm sys}\times10^{-9},\\
 &\frac{\mathcal{B}(\eta'\rightarrow\gamma(\mu^+\mu^-))}{\mathcal{B}(\eta\rightarrow\gamma\gamma)}=1.761(7)_{\rm stat}(2)_{\rm sys}\times10^{-9},
 \end{align}
With this branching ratio, about 1000 $\eta$ events would be produced at REDTOP with their projected luminosity, but this number should be reduced by detector efficiencies and cuts. 
While no specific experiment has yet been proposed, a similar search could be undertaken in future high-intensity $K_L$ beams being considered at CERN and J-PARC, where the rate has been computed to NLO and including modern experimental form factors~\cite{Ji:2017lyh}: 
 \begin{align}
  &\frac{\mathcal{B}(K_L\rightarrow\gamma(\mu^+\mu^-))}{\mathcal{B}(\eta\rightarrow\gamma\gamma)}\approx1.26(2)_{\rm model}\times10^{-9}.\nonumber
\end{align}


\begin{acknowledgement}
The authors would like to thank N. Raman for his unpublished calculations.  HL is supported by the U.S. Department of Energy under Contract No. DE-FG02-93ER-40762.  YJ acknowledges the Deutsche Forschungsgemeinschaft for support under grant BR 2021/7-1.
\end{acknowledgement}

\bibliographystyle{woc.bst}
\bibliography{/home/hlamm/wise}

\begin{thebibliography}{50}

\bibitem{PhysRevD.73.072003}
G.~Bennett et~al. (Muon G-2 Collaboration), Phys. Rev. \textbf{D73}, 072003
  (2006), \texttt{hep-ex/0602035}

\bibitem{Grange:2015fou}
J.~Grange et~al. (Muon g-2) (2015), \texttt{1501.06858}

\bibitem{Saito:2012zz}
N.~Saito (J-PARC g-'2/EDM), AIP Conf. Proc. \textbf{1467}, 45 (2012)

\bibitem{Antognini:1900ns}
A.~Antognini et~al., Science \textbf{339}, 417 (2013)

\bibitem{Pohl1:2016xoo}
R.~Pohl et~al. (CREMA), Science \textbf{353}, 669 (2016)

\bibitem{Beyer79}
A.~Beyer, L.~Maisenbacher, A.~Matveev, R.~Pohl, K.~Khabarova, A.~Grinin,
  T.~Lamour, D.C. Yost, T.W. H{\"a}nsch, N.~Kolachevsky et~al., Science
  \textbf{358}, 79 (2017)

\bibitem{Aaij:2014ora}
R.~Aaij et~al. (LHCb collaboration), Phys. Rev. Lett. \textbf{113}, 151601
  (2014), \texttt{1406.6482}

\bibitem{Aaij:2015yra}
R.~Aaij et~al. (LHCb), Phys. Rev. Lett. \textbf{115}, 111803 (2015), [Addendum:
  Phys. Rev. Lett.115,no.15,159901(2015)], \texttt{1506.08614}

\bibitem{Ciezarek:2017yzh}
G.~Ciezarek, M.~Franco~Sevilla, B.~Hamilton, R.~Kowalewski, T.~Kuhr, V.~Luth,
  Y.~Sato (2017), \texttt{1703.01766}

\bibitem{Aaij:2017tyk}
R.~Aaij et~al. (LHCb) (2017), \texttt{1711.05623}

\bibitem{Aaij:2017vbb}
R.~Aaij et~al. (LHCb), JHEP \textbf{08}, 055 (2017), \texttt{1705.05802}

\bibitem{TuckerSmith:2010ra}
D.~Tucker-Smith, I.~Yavin, Phys. Rev. \textbf{D83}, 101702 (2011),
  \texttt{1011.4922}

\bibitem{PhysRevD.91.073008}
H.~Lamm, Phys. Rev. D \textbf{91}, 073008 (2015)

\bibitem{Lamm:2015gka}
H.~Lamm, Phys. Rev. \textbf{D92}, 055007 (2015), \texttt{1505.00057}

\bibitem{Lamm:2016jim}
H.~Lamm, Phys. Rev. \textbf{D94}, 115007 (2016), \texttt{1609.07520}

\bibitem{Hughes:1971}
V.~Hughes, B.~Maglic, Bull. Am. Phys. Soc. \textbf{16}, 65 (1971)

\bibitem{baier1962bimuonium}
V.N. Baier, V.S. Synakh, JETP \textbf{14}, 1122 (1962)

\bibitem{budini1961reactions}
P.~Budini, Tech. rep., CM-P00056754 (1961)

\bibitem{Karshenboim:2005iy}
S.G. Karshenboim, Phys. Rept. \textbf{422}, 1 (2005), \texttt{hep-ph/0509010}

\bibitem{Lamm:2016vtf}
H.~Lamm, Phys. Rev. \textbf{A95}, 012505 (2017), \texttt{1611.04258}

\bibitem{Adkins:2014dva}
G.S. Adkins, R.N. Fell, Phys. Rev. \textbf{A89}, 052518 (2014),
  \texttt{1402.7040}

\bibitem{PhysRevA.94.032507}
Y.~Ji, H.~Lamm, Phys. Rev. A \textbf{94}, 032507 (2016)

\bibitem{Jentschura:1997tv}
U.~Jentschura, G.~Soff, V.~Ivanov, S.G. Karshenboim, Phys. Rev. \textbf{A56},
  4483 (1997), \texttt{physics/9706026}

\bibitem{Lamm:2015fia}
H.~Lamm, \emph{{True muonium: the atom that has it all}}, in \emph{{Twelfth
  International Conference on the Intersections of Particle and Nuclear
  Physics, Vail, CO, USA, May 19-24, 2015}} (2015), \texttt{1509.09306},
  \urlstyle{tt}\url{https://inspirehep.net/record/1395477/files/arXiv:1509.09306.pdf}

\bibitem{Adkins:2015jia}
G.S. Adkins, M.~Kim, C.~Parsons, R.N. Fell, Phys. Rev. Lett. \textbf{115},
  233401 (2015), \texttt{1507.07841}

\bibitem{Lamm:2017lib}
Y.~Ji, H.~Lamm (2017), \texttt{1701.04362}

\bibitem{Naveen}
H.~Lamm, N.~Raman, in prep.  (2017)

\bibitem{Robiscoe:1965zz}
R.T. Robiscoe, Phys. Rev. \textbf{138}, A22 (1965)

\bibitem{Brodsky:2009gx}
S.J. Brodsky, R.F. Lebed, Phys. Rev. Lett. \textbf{102}, 213401 (2009),
  \texttt{0904.2225}

\bibitem{Nemenov:2001vp}
L.L. Nemenov, V.D. Ovsyannikov, Phys. Lett. \textbf{B514}, 247 (2001)

\bibitem{Celentano:2014wya}
A.~Celentano (HPS), J. Phys. Conf. Ser. \textbf{556}, 012064 (2014)

\bibitem{dirac}
P.~Chliapnikov, DIRAC-NOTE-2014-05  (2014)

\bibitem{Bogomyagkov:2017uul}
A.~Bogomyagkov, V.~Druzhinin, E.~Levichev, A.~Milstein, S.~Sinyatkin (2017),
  \texttt{1708.05819}

\bibitem{Badertscher:2006fm}
A.~Badertscher, P.~Crivelli, W.~Fetscher, U.~Gendotti, S.~Gninenko, V.~Postoev,
  A.~Rubbia, V.~Samoylenko, D.~Sillou, Phys. Rev. \textbf{D75}, 032004 (2007),
  \texttt{hep-ex/0609059}

\bibitem{Bilenky:1969zd}
S.M. Bilenky, V.H. Nguyen, L.L. Nemenov, F.G. Tkebuchava, Yad. Fiz.
  \textbf{10}, 812 (1969)

\bibitem{Holvik:1986ty}
E.~Holvik, H.A. Olsen, Phys. Rev. \textbf{D35}, 2124 (1987)

\bibitem{ArteagaRomero:2000yh}
N.~Arteaga-Romero, C.~Carimalo, V.~Serbo, Phys. Rev. \textbf{A62}, 032501
  (2000), \texttt{hep-ph/0001278}

\bibitem{baier1961bimuonium}
V.~Baier, V.~Synakh, Zhur. Eksptl'. i Teoret. Fiz. \textbf{41} (1961)

\bibitem{Moffat:1975uw}
J.~Moffat, Phys. Rev. Lett. \textbf{35}, 1605 (1975)

\bibitem{Nemenov:1972ph}
L.~Nemenov, Yad. Fiz. \textbf{15}, 1047 (1972)

\bibitem{Kozlov:1987ey}
G.~Kozlov, Sov. J. Nucl. Phys. \textbf{48}, 167 (1988)

\bibitem{Malenfant:1987tm}
J.~Malenfant, Phys. Rev. \textbf{D36}, 863 (1987)

\bibitem{Ginzburg:1998df}
I.~Ginzburg, U.~Jentschura, S.G. Karshenboim, F.~Krauss, V.~Serbo et~al., Phys.
  Rev. \textbf{C58}, 3565 (1998), \texttt{hep-ph/9805375}

\bibitem{Chen:2012ci}
Y.~Chen, P.~Zhuang (2012), \texttt{1204.4389}

\bibitem{Banburski:2012tk}
A.~Banburski, P.~Schuster, Phys. Rev. \textbf{D86}, 093007 (2012),
  \texttt{1206.3961}

\bibitem{Ellis:2015eea}
S.~Ellis, J.~Bland-Hawthorn (2015), \texttt{1501.07281}

\bibitem{Ellis:2017kzh}
S.C. Ellis, J.~Bland-Hawthorn (2017), \texttt{1712.02022}

\bibitem{Benelli:2012bw}
A.~Benelli (DIRAC Collaboration), EPJ Web Conf. \textbf{37}, 01011 (2012)

\bibitem{Gatto:2016rae}
C.~Gatto, B.~Fabela~Enriquez, M.I. Pedraza~Morales (REDTOP), PoS
  \textbf{ICHEP2016}, 812 (2016)

\bibitem{Ji:2017lyh}
Y.~Ji, H.~Lamm (2017), \texttt{1706.04986}

\end{thebibliography}

\end{document}